\documentclass{SciPost}

\hypersetup{
    colorlinks,
    linkcolor={red!50!black},
    citecolor={blue!50!black},
    urlcolor={blue!80!black}
}

\usepackage[bitstream-charter]{mathdesign}
\usepackage{amsmath}
\usepackage{amssymb}
\usepackage{float}
\usepackage{array}
\usepackage{booktabs}
\usepackage{caption}
\usepackage{subcaption}
\usepackage{comment}
\usepackage{setspace}
\usepackage{url}
\DeclareCaptionType{equ}[][]

\DeclareSymbolFont{usualmathcal}{OMS}{cmsy}{m}{n}
\DeclareSymbolFontAlphabet{\mathcal}{usualmathcal}

\fancypagestyle{SPstyle}{
\fancyhf{}
\lhead{\colorbox{scipostblue}{\bf \color{white} ~SciPost Physics Proceedings }}
\rhead{{\bf \color{scipostdeepblue} ~Submission }}

\fancyfoot[C]{\textbf{\thepage}}
}

\begin{document}

\pagestyle{SPstyle}

\begin{center}{\Large \textbf{\color{scipostdeepblue}{
Production and propagation of secondary antideuteron in the Galaxy\\
}}}\end{center}

\begin{center}\textbf{
Luis Fernando Galicia Cruztitla\textsuperscript{$\dagger$}
 and
Diego Mauricio Gómez Coral\textsuperscript{$\star$}
}\end{center}

\begin{center}
 Instituto de Física, Universidad Nacional Autónoma de México, Circuito de la Investigación Científica, Ciudad Universitaria, CDMX, 04510, Mexico
\\
[\baselineskip]
$\dagger$ \href{mailto:email2}{\small luisfergac@ciencias.unam.mx}\,,\quad
$\star$ \href{mailto:email1}{\small dgomezco@fisica.unam.mx}
\end{center}

\definecolor{palegray}{gray}{0.95}
\begin{center}
\colorbox{palegray}{
  \begin{tabular}{rr}
  \begin{minipage}{0.36\textwidth}
    \includegraphics[width=55mm]{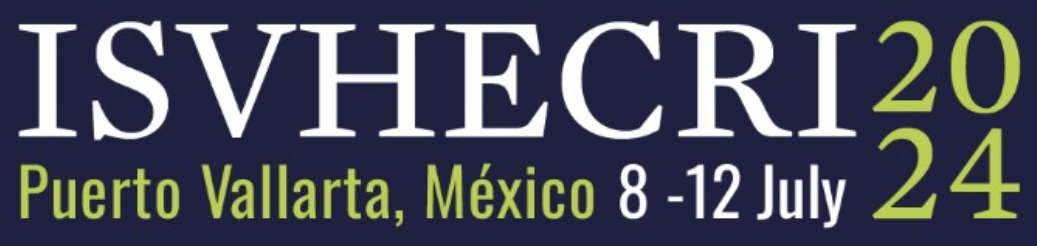}
  \end{minipage}
  &
  \begin{minipage}{0.55\textwidth}
    \begin{center} \hspace{5pt}
    {\it 22nd International Symposium on Very High \\Energy Cosmic Ray Interactions (ISVHECRI 2024)} \\
    {\it Puerto Vallarta, Mexico, 8-12 July 2024} \\
    \doi{10.21468/SciPostPhysProc.?}\\
    \end{center}
  \end{minipage}
\end{tabular}
}
\end{center}

\section*{\color{scipostdeepblue}{Abstract}}
\textbf{\boldmath{%
This work reviews the current state of the antideuteron ($\bar{d}$) production cross-sections in cosmic ray interactions and its uncertainties, considering the coalescence model and measurements in accelerator experiments. These cross-sections have been included in a simulation of cosmic rays propagation in the Galaxy using \textbf{{\tt GALPROP\,v.57}}, with updated parameters of the diffusive reacceleration model. An estimation of the expected antideuteron flux at Earth is presented.
}}

\vspace{\baselineskip}

\noindent\textcolor{white!90!black}{%
\fbox{\parbox{0.975\linewidth}{%
\textcolor{white!40!black}{\begin{tabular}{lr}%
  \begin{minipage}{0.6\textwidth}%
    {\small Copyright attribution to authors. \newline
    This work is a submission to SciPost Phys. Proc. \newline
    License information to appear upon publication. \newline
    Publication information to appear upon publication.}
  \end{minipage} & \begin{minipage}{0.4\textwidth}
    {\small Received Date \newline Accepted Date \newline Published Date}%
  \end{minipage}
\end{tabular}}
}}
}


\vspace{10pt}
\noindent\rule{\textwidth}{1pt}
\tableofcontents
\noindent\rule{\textwidth}{1pt}
\vspace{10pt}


\section{Introduction}
\label{sec:intro}
Cosmic Rays (\textbf{CRs}) produced in stars and accelerated into the Interstellar Medium (\textbf{ISM}) in supernova remnants (\textbf{SNRs}) are called \textit{primary cosmic rays}. Composed mostly of protons (p) and Helium (He), these primary particles interact with the \textbf{ISM} through inelastic collisions producing \textit{secondary cosmic rays} \cite{Grupen_2020}. In a Universe dominated by matter, and within the Standard Model of particles (\textbf{SM}), the production of antimatter could only be possible through the nuclear interactions of \textbf{CRs} with the \textbf{ISM}, i.e. antimatter is expected to be of \textit{secondary} origin. However, Dark Matter (\textbf{DM}) models predict $\bar{d}$ production by annihilation or decay of \textbf{DM} particles in the Galaxy\,\cite{Donato_2008}. This additional primary $\bar{d}$ production would be observed as an excess to the secondary component at energies below 1\,GeV\,\cite{Doetinchem_2020}. Interestingly, the Alpha Magnetic Spectrometer (AMS-02) detector onboard the International Space Station (\textbf{ISS}) has reported 7 $\bar{d}$ candidates in the energy region above 1\,GeV\,\cite{Ting_2023}. Although a \textbf{DM} origin of these $\bar{d}$ seems to be in tension with the observations, a clear interpretation of the results requires a good estimation of the secondary component. Therefore, in this work a review of the $\bar{d}$ production cross sections within the coalescence model, and an estimation of the flux after propagation in the Galaxy are presented.


\section{Transport of galactic CRs}
\label{sec: transport}

Charged \textbf{CRs} particles propagate in the Galaxy following a diffusive process due to perturbations in the magnitude and direction of the magnetic fields of the different regions of the Galaxy. As a consequence, \textbf{CRs} confined time in the galactic volume is on the order of mega-years\cite{Blasi_2023}. The dynamics of the propagation of \textbf{CRs} in the Galaxy is modeled by the \textit{transport equation} \cite{_erk_nyt__2022}:

\begin{eqnarray}    \label{eq:treq}
    \frac{\partial\psi(\bar{r}, p, t)}{\partial t} = Q(\bar{r}, p) + \nabla \cdot [D_{xx}\nabla\psi - \bar{V}\psi] + \frac{\partial}{\partial p} \left[ p^{2}D_{pp}\frac{\partial}{\partial p}\frac{\psi}{p^{2}} \right] - \nonumber\\ 
    \frac{\partial}{\partial p} \left[\frac{dp}{dt}\psi - \frac{1}{3}p\nabla \cdot \bar{V}\psi\right] - \frac{\psi}{\tau_{f}} - \frac{\psi}{\tau_{d}}.
    \end{eqnarray}

On the left side, we have the time variation of the $\psi$ function that represents the density of particles per unit of momentum. On the right side, the first term ($Q(\bar{r}, p)$) represents the sources of \textbf{CRs} such as \textbf{SNRs} or secondary production by nuclear interactions. The second term corresponds to the diffusion and convective part, where $D_{xx}$ is the diffusion coefficient and $V$ is the convection velocity. The third term refers to the diffusive reacceleration process, with a diffusion coefficient in the momentum space $D_{pp}$. The fourth term describes energy loss processes, in particular, momentum loss due to interactions with the \textbf{ISM} and losses due to non-uniform convection. Finally, the last two terms indicate losses by fragmentation or annihilation and losses by radioactive decay respectively.

To simulate the propagation process of $\bar{d}$ in the Galaxy, the open source code \textbf{{\tt GALPROP v.57}}\,\cite{Porter_2022} is used. \textbf{{\tt GALPROP}} solves the transport equation numerically, and introduces current astrophysical models and data to make the simulation more realistic. For the present study, a two-dimensional diffusion halo with cylindrical symmetry is considered. The parameters on the transport equation (including the size of the Galaxy, and diffusion coefficient, among others), were set according to Boschini\,\textit{et al.}\,\cite{Boschini_2020}.

\section{Secondary antideuteron production}
\subsection{Coalescence model}
\label{sec:sec_dbar}
The coalescence model postulates that an $\bar{d}$ can be formed if an antiproton ($\bar{p}$) and an antineutron ($\bar{n}$) produced after an interaction are close in phase space within a volume determined by a radius in momentum $p_0$\,\cite{Butler_1963, Csernai_1986}, known as the coalescence momentum. In this way, a primary accelerated p or He, with enough energy, interacting with a p or He of the \textbf{ISM}, might produce a sequence of $\bar{p}$ and $\bar{n}$ that could coalesce to form $\bar{d}$ or increasingly heavier antinuclei. 

As a first analytical approach, the distribution of the $\bar{p}$-$\bar{n}$ pair momentum can be seen as the product of two independent isotropic distributions, without correlation\,\cite{Chardonnet_1997}. In this way, the $\bar{d}$ spectrum is given by

\begin{equation}
    \gamma_{\bar{d}}\frac{\partial^{3}N_{\bar{d}}}{dp_{\bar{d}}^{3}} = \frac{4\pi}{3} p_{0}^{3} \left(\gamma_{\bar{p}} \frac{\partial^{3}N_{\bar{p}}}{dp_{\bar{p}}^{3}}\right) \left(\gamma_{\bar{n}} \frac{\partial^{3}N_{\bar{n}}}{dp_{\bar{n}}^{3}}\right),
\end{equation}

\noindent where $\frac{\partial^{3}N_{\bar{}}}{dp_{\bar{}}^{3}}$ represents the differential yield per event of particle ($\bar{p}$, $\bar{n}$ and $\bar{d}$) in terms of momentum, $\gamma$ is the Lorentz factor for each particle and $p_{0}$ is the coalescence momentum. However, the production of antinuclei pairs in nuclear collisions is not independent or isotropic, i.e. they are correlated\,\cite{Gomez_Coral_2018}. To take these correlations into account, the coalescence mechanism to form $\bar{d}$ is simulated on an event-by-event basis, using $\bar{p}$-$\bar{n}$ pairs produced in p-p, p-He, He-p, and He-He collisions with Monte Carlo (\textbf{MC}) generators. Collisions simulated with \textbf{MC} generators follow interaction models that already contain correlations between products\,\cite{Kachelriess_2015}. To generate a new $\bar{d}$ from any $\bar{p}$-$\bar{n}$ pair, the center of mass momentum of the pair $p_{cm}$ should be less than the coalescence momentum, i.e. $p_{cm} < p_{0}$. The value of $p_{0}$ is chosen in a way the final $\bar{d}$ differential cross section is in agreement to accelerator measurements\,\cite{Gomez_Coral_2018}. On the other hand, the condition on the $\bar{p}$-$\bar{n}$ pair separation boundary ($\Delta x$) is enforced by allowing only $\bar{d}$ that are generated within a radius of $\sim$\,2\,fm to coalesce. Here, \textbf{CRs} collisions and $\bar{d}$ production simulated with \textbf{{\tt EPOS-LHC}} generator and a coalescence afterburner by Shukla \textit{et al.}\,\cite{Shukla_2020} and Gomez-Coral \textit{et al.}\,\cite{Gomez_Coral_2018} were used. 

\subsection{Secondary source term} 

The secondary source term $Q$ (see Eq. \ref{eq:treq}) for $\bar{d}$ production by \textbf{CRs} collisions can be written as follows:

\begin{equation} \label{eq:srcterm}
    Q_{\bar{d}}^{sec}(E_{kin}^{\bar{d}}, \mathbf{r}) = \sum_{i\in\{p, He, \bar{p}\}} \sum_{j\in\{p, He\}} 4\pi n_{j}(\mathbf{r}) \int_{E^{(i,j)}_{kin(\bar{d})}}^{\infty} \frac{d \sigma_{i,j}(E_{kin(i)}, E_{kin}^{\bar{d}})}{dE^{\bar{d}}_{kin}} \Phi_{i}(E_{kin(i)}, \mathbf{r})\,dE_{kin},
\end{equation}

\noindent where $n_{j}$ is the particle density of the \textbf{ISM}, $E_{kin(i)}$ denote the kinetic energy per nucleon of the incident particles, $\Phi_{i}$ the incident flux of \textbf{CRs} (p and He) and $\left(\frac{d\sigma_{prod}}{dE_{kin}^{\bar{d}}}\right)_{i,j}$ corresponds to the differential cross section for $\bar{d}$ with $E_{kin}^{\bar{d}}$ the kinetic energy per nucleon for $\bar{d}$\,\cite{Ibarra_2013}. The secondary source term convolves the $\bar{d}$ differential production cross section obtained using \textbf{{\tt EPOS-LHC}} and the coalescence model\,\cite{Gomez_Coral_2020} (see Sec.\,\ref{sec:sec_dbar}).

\section{Results}
\subsection{Differential cross section production and source term for $\bar{d}$}
\begin{figure}[!h]
\begin{subfigure}[h]{0.45\linewidth}
\includegraphics[width=\linewidth]{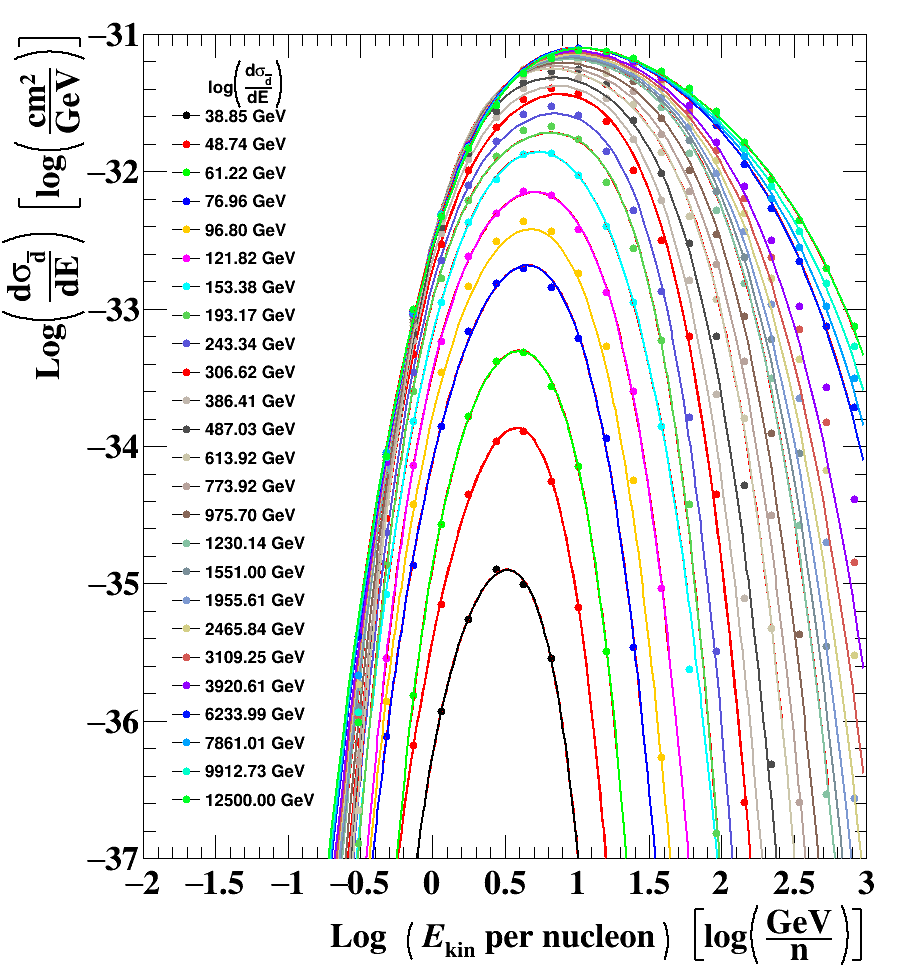}
\caption{}
\end{subfigure}
\hfill
\begin{subfigure}[h]{0.45\linewidth}
\includegraphics[width=\linewidth]{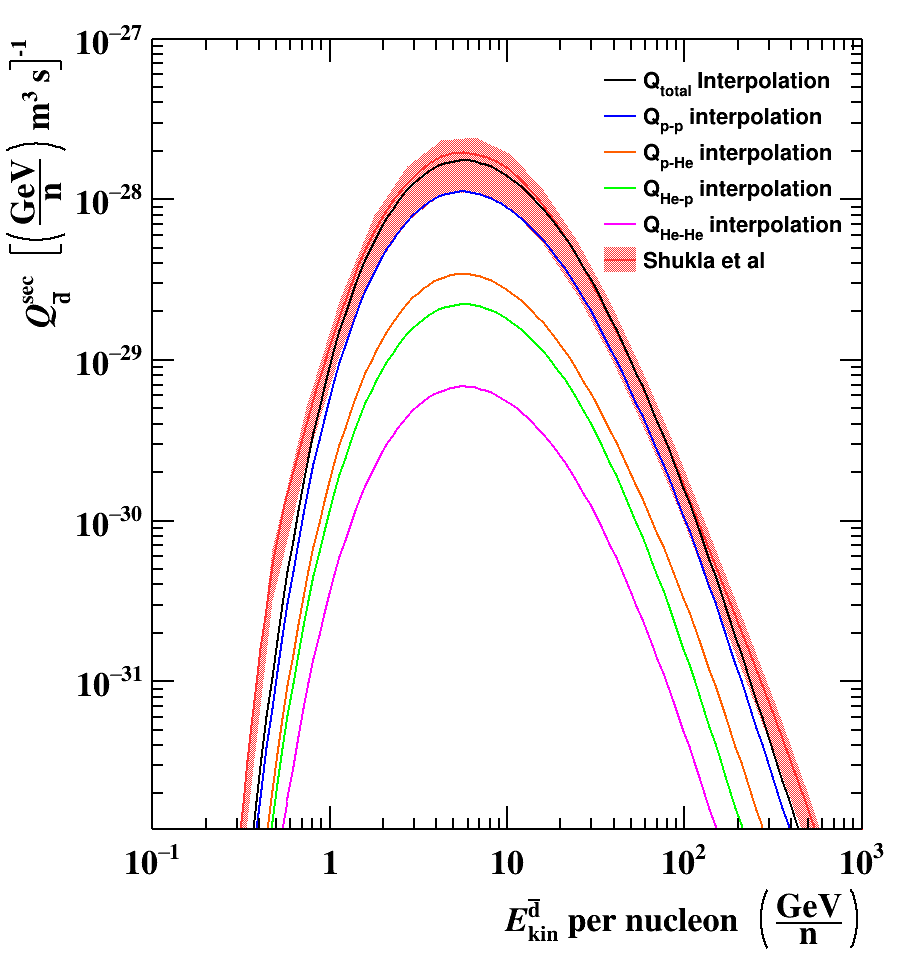}
\caption{}
\end{subfigure}%
\caption{(a): $\bar{d}$ production differential cross section ($\frac{d\sigma_{\bar{d}}}{dE}$), as a function of $\bar{d}$ kinetic energy per nucleon in log-log scale, for different projectile energies. (b): Total secondary $\bar{d}$ source term (black line) and contributions from all types of \textbf{CRs} collisions (color lines) calculated with a parametrization function and compared to \cite{Shukla_2020} (red band).}
\label{subfig:log_cs}
\end{figure}

From the limited discrete data generated with \textbf{{\tt EPOS-LHC}} and the coalescence simulation of $\bar{d}$ published by Shukla \textit{et al}\,\cite{Shukla_2020} (color points in Fig.\ref{subfig:log_cs} (a)), a polynomial fit was performed over a continuous interval of $\bar{d}$ kinetic energy per nucleon, for every set of data points per projectile kinetic energy (color lines in Fig.\ref{subfig:log_cs} (a)). Then, the parameters obtained from this fit were interpolated as a function of the projectile kinetic energy per nucleon, to get a complete parametrization of the $\bar{d}$ production cross sections. 

\sloppy Then, the secondary source term was estimated using the $\bar{d}$ production cross section parametrization in Eq.\,\ref{eq:srcterm}, and was compared to \cite{Shukla_2020}. The results are shown in Fig.\,\ref{subfig:log_cs} (b), where can be seen that both results (solid black line and red solid line) are in agreement for most of the $\bar{d}$ energy range. 

\subsection{Expected $\bar{d}$ flux}

\begin{figure}[!h]
    \centering
    \includegraphics[width=0.5\linewidth]{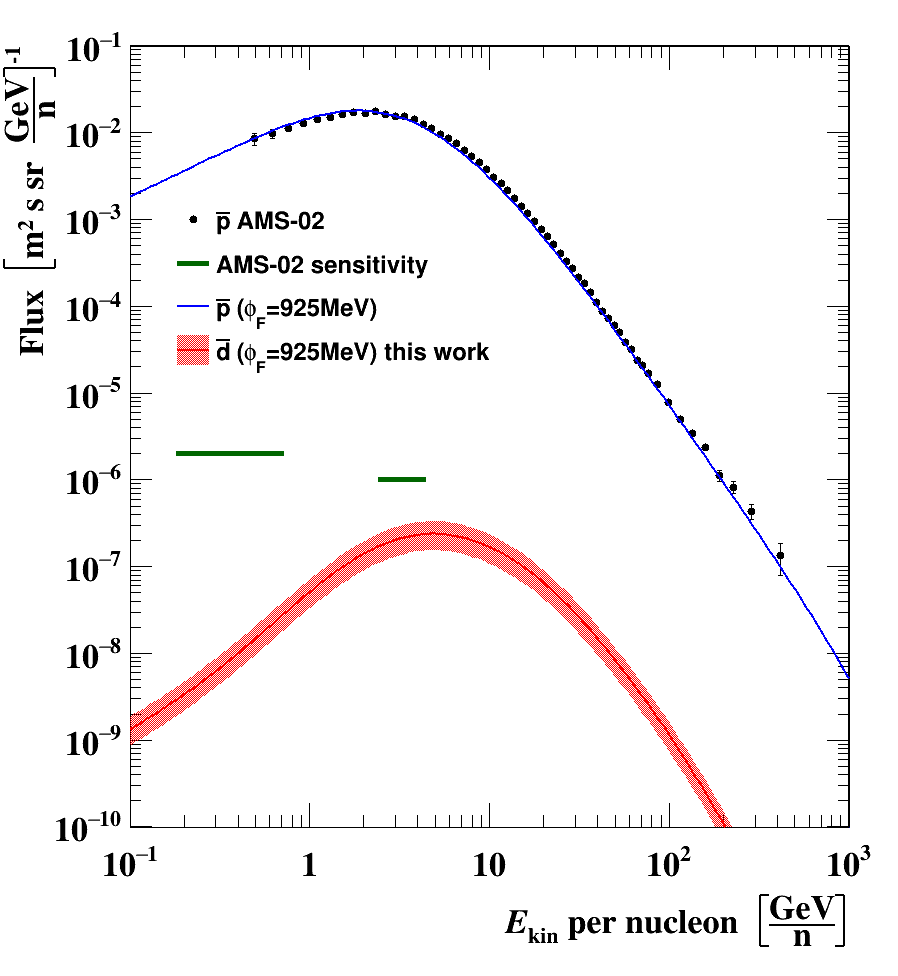}
    \caption{Estimated secondary $\bar{d}$ flux with \textbf{{\tt GALPROP\,v.57}} (red line) and its uncertainties associated with cross sections (red band). For comparison, the $\bar{d}$ AMS-02 sensitivity is shown as green lines\,\cite{Aramaki_2016}. The modulated $\bar{p}$ flux calculated with \textbf{{\tt GALPROP\,v.57}} (blue line) is compared to $\bar{p}$ AMS-02 data\,\cite{Aguilar_2021} (black points).}
    \label{fig:dbar_flux}
\end{figure}

Finally, the $\bar{d}$ production cross section parametrization was implemented in \textbf{{\tt GALPROP\,v.57}} (Sec.\,\ref{sec: transport}) where $\bar{d}$ were propagated using updated propagation parameters. Solar modulation was modeled using the Force Field approximation\,\cite{Gleeson_1968}. $\phi_F$ was obtained from a fit of the model to AMS-02 $\bar{p}$ measurements\,\cite{Aguilar_2021}. The resulting Top-Of-Atmosphere $\mathrm{TOA}$ fluxes as a function of kinetic energy per nucleon for $\bar{p}$ (blue solid line) and $\bar{d}$ (red line) are shown in Fig.\,\ref{fig:dbar_flux} along with AMS-02 $\bar{p}$ data \cite{Aguilar_2021} (black points) and AMS-02 $\bar{d}$ sensitivity \cite{Aramaki_2016} (green lines), considering the projection with a superconducting-magnet configuration. The red band in Fig.\,\ref{fig:dbar_flux} represents the $\bar{d}$ cross section uncertainty estimated in\,\cite{Shukla_2020} as $35\%$ above and $45\%$ below the expected flux.

\section{Conclusions}
The expected secondary $\bar{d}$ flux was estimated by performing a parametrization of the $\bar{d}$ production cross sections as a function of $\bar{d}$ kinetic energy and projectile energy, using \textbf{MC} data generated by\,\cite{Shukla_2020}. To propagate $\bar{d}$ in the Galaxy, \textbf{{\tt GALPROP\,v.57}} was used along with the latest propagation parameters from\,\cite{Boschini_2020}. The final results are presented in Fig.\,\ref{fig:dbar_flux}, where it can be seen the AMS-02 sensitivity reported in\,\cite{Aramaki_2016} is above the calculated $\bar{d}$ flux by a factor of 3. A similar result is obtained when an AMS-02 sensitivity, with a permanent-magnet configuration for 11 years, is considered\,\cite{Weng_2023}. Previous calculations are consistent with this result\,\cite{Blum_2017, Korsmeier_2018}. This suggests the $\bar{d}$ candidates reported by AMS-02\,\cite{Ting_2023} would not come from a secondary origin. However, measurements in $\bar{d}$ production cross sections for projectile energies between 100 to 200\,GeV are necessary to corroborate a coalescence momentum dependence with projectile energy\,\cite{Gomez_Coral_2018}. A significantly higher $\bar{d}$ flux can be obtained by considering a constant $p_0$\,\cite{Luque_2024}. Interestingly, a recent AMS-02 sensitivity projection for 2030 with a planned detector upgrade is close to the maximum of our secondary $\bar{d}$ flux estimation\,\cite{Weng_2023} (Fig.\,\ref{fig:dbar_flux}), suggesting that AMS-02 would be practically background-free for possible antideuteron observation coming from dark matter or new astrophysical production mechanisms.



 



\section*{Acknowledgements}
The authors thank Anirvan Shukla for sharing the \textbf{MC}-generated data. This work was supported by UNAM-PAPIIT IA101624 and by CONAHCYT under grant CBF2023-2024-118. LFGC thanks the support from CONAHCYT-[CVU-937792] scholarship and PAEP-UNAM.







\bibliography{proceedings_isvhecri2024}



\end{document}